# Integrated Sensor System to Control the Temperature Effects and the Hysteresis on Adaptive Fluid-Membrane Piezoelectric Lenses

Hitesh G. B. Gowda and Ulrike Wallrabe

Laboratory for Microactuators, Department for Microsystems Engineering - IMTEK, University of Freiburg, Germany.

hitesh.gowda@imtek.uni-freiburg.de; wallrabe@imtek.uni-freiburg.de

## ABSTRACT

We present in this paper an integrated sensor system for closed-loop control of the temperature effects and the hysteresis on the refractive power of an adaptive fluid-membrane piezoelectric lens. The piezoelectric hysteresis and the fluid thermal expansion contribute to a nonlinear response of the lens refractive power defined as a function of the actuation electric field. Hence, a pressure sensor and a temperature sensor are integrated inside the lens to monitor and define the lens refractive power as a function of both the internal fluid pressure and the temperature, thus, allowing for the closed-loop control of the refractive power. The adaptive lens has a refractive power range varying from -16 $m^{-1}$ to +17 $m^{-1}$ at 25°C and from -15 $m^{-1}$ to +28 $m^{-1}$ at 75°C.

## 1. INTRODUCTION

The adaptive optics technology was initially used in astronomical applications to modify the phase of the incoming light by using a deformable mirror [1]. Eventually, the adaptive optics technology was used in microscopes for correcting aberrations [2], optical communication systems for coupling light [3], and in optical imaging systems for varying focal power [4]. In the adaptive optics technology, the surface of the optical component is deformed by a microactuator to modify the characteristics of the incident light. The microactuator contributes to a smaller size, lower power consumption, and a faster response system [5]. An imaging system implemented with an adaptive optics lens eliminates the conventional mechanical movement of lenses to focus an image and contributes to the faster response and longer lifespan of the system. The focal power of the adaptive lens, which is dependent on the surface curvature, is varied by deforming the flexible surface of the lens with the use of a microactuator. One such adaptive lens utilizing a piezoelectric microactuator to deform a flexible fluid-membrane interface was developed in the Laboratory for Microactuators, Department of Microsystems Engineering- IMTEK, University of Freiburg, Germany.

The developed adaptive lens working principle relies on the interaction between the integrated annular piezoelectric actuator and the fluid chamber bounded by a flexible membrane [6]. The piezoelectric actuator deforms the fluid chamber and varies the internal fluid pressure. The varied fluid pressure acts on the flexible membrane and deforms it into an aspherical surface with varied surface curvature. Hence, a change in the internal fluid pressure results in a change in the refractive power of the lens. The refractive power defined as a function of the actuation electric field at the piezo exhibits a non-linear hysteresis. Hence, Draheim et al. integrated a relative pressure sensor inside the lens to define and control the refractive power as a function of the internal fluid pressure [7]. However, at higher temperatures, the thermal expansion of the fluid affects the fluid pressure as well and results in a deviation of the defined refractive power. In this paper, we present a further integration of a temperature sensor inside the lens to account for the temperature effects and to define the refractive power as a function of both the internal fluid pressure and the temperature. Besides, we have adapted our process to mount the lens chamber directly on a PCB substrate.

## 2. DESIGN

The adaptive lens presented by Draheim et al. [6] consists of a piezoelectric actuator, a fluid chamber, and a flexible transparent PDMS membrane to form an active lens chamber. The piezoelectric actuator is manufactured by cutting the piezoelectric ceramics into annular discs using a UV laser and gluing them using a two-component epoxy in antiparallel polarization configuration. Further, the transparent membrane and the fluid chamber are molded onto the actuator using micromolding techniques. In our new PCB based design, the sensors and the electrical connectors are integrated on the PCB, which is also equipped with a central glass window. The PCB is commercially fabricated and is modified using a CNC machine to form slots for the lens chamber, the glass window, and the pressure sensor, as shown in Figure 1 (a) and (b). A PTC temperature sensor is soldered directly on the PCB, and a membrane-based relative pressure sensor is glued inside the previously machined slot and wire bonded for electrical connections (Figure 1 (b)). The separately fabricated lens chamber is glued with PDMS on the PCB and primed with an optical oil. The adaptive lens has a clear aperture of 10 mm at an outer diameter of 20 mm, as shown in Figure 1 (d).

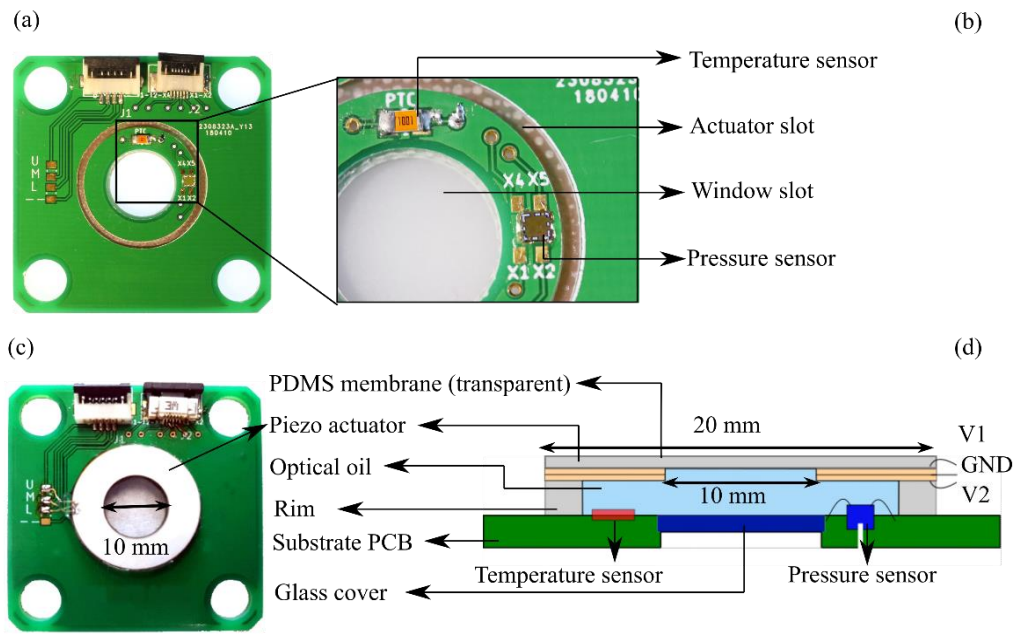

Figure 1 (a) Integrated PCB with sensors and connectors, (b) enlarged view to show temperature and pressure sensor, (c) adaptive lens mounted on the PCB substrate, and (d) 2D cross-sectional view of the adaptive lens showing the integrated temperature and pressure sensor inside the fluid chamber.

In our bimorph actuator, the piezoceramics are configured with anti-parallel polarization. Under the application of an electric field on the piezoelectric bimorph, one piezo expands and the other contracts resulting in a bending deflection. The actuator bending deflection will deform the fluid chamber and vary the internal fluid pressure. The varied fluid pressure acts on the flexible PDMS membrane and deforms it in the opposite direction to that of the actuator deflection. Due to the circular design of the adaptive lens, the membrane deforms into an aspherical surface. By varying the magnitude and direction of the applied electric field on the piezoelectric actuator, a plano-convex lens (Figure 2 (a)), or a plano-concave lens (Figure 2 (b)) with a variable refractive power can be achieved.

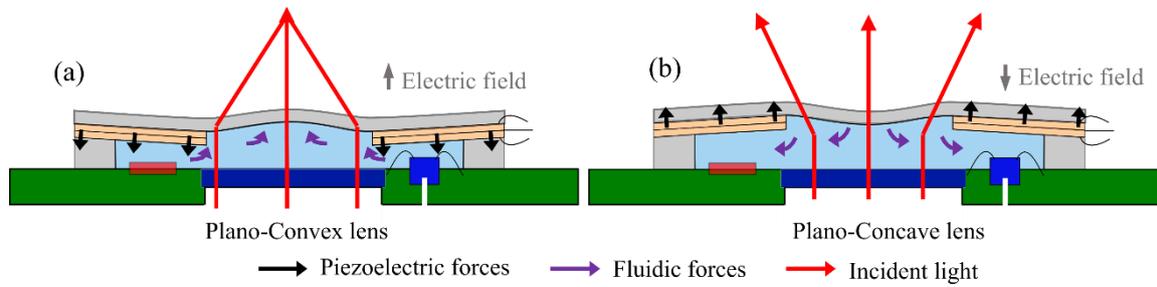

Figure 2 Adaptive lens under applied electric field at the piezo to form (a) a plano-convex lens or (b) a plano-concave lens.

## 3. CHARACTERIZATION

The adaptive lens was characterized for the refractive power at different actuation levels and higher temperatures. For the characterization of the adaptive lens at higher temperatures, the adaptive lens (Figure 3 (b)) was mounted on a resistive heater (Figure 3 (a)) and heated to required temperatures. The membrane surface profile was measured using a profilometer and a confocal displacement sensor, providing a resolution of 110 nm (Figure 3 (c)). A sinusoidal electric field of 1.3 kV/mm at 1 Hz was applied to the piezo actuator through an electrical driver. The electrical driver was used to divide the electric field for the two piezoceramics in the bimorph and also to limit the negative electric field to 1/3 of the piezoelectric coercive field to avoid depolarization [8]. The refractive power was calculated from the radius of curvature of the measured membrane profile. The characterization was repeated with the adaptive lens heated to higher temperatures.

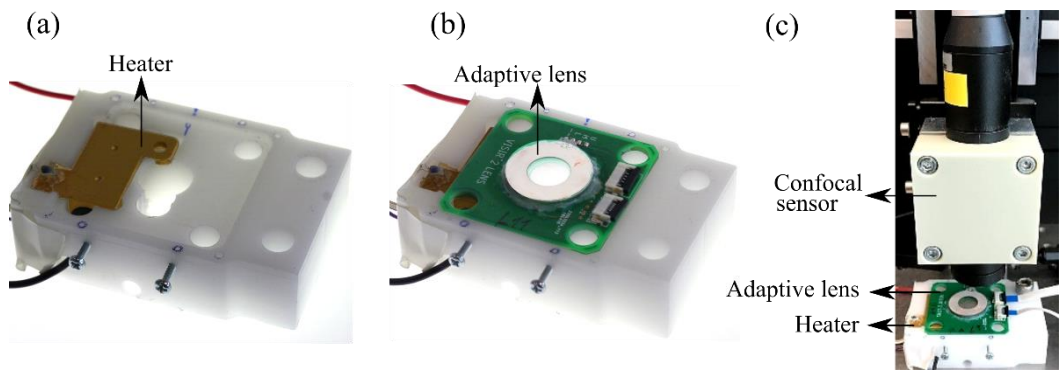

Figure 3 The characterization setup showing (a) heating element, (b) adaptive lens mounted on the heater, and (c) confocal sensor to measure membrane surface.

Figure 4 (a) shows the refractive power as a function of the applied electric fields at different temperatures. A hysteresis, as well as a temperature drift, are obvious. During the characterization, the outputs from the pressure sensor and the temperature sensor were measured synchronously with the membrane deformation. The measured sensor outputs and membrane deformation were used to define the refractive power as a function of both the internal fluid pressure and the temperature, as shown in Figure 4 (b). The integrated sensors were used to compensate for the hysteresis and the temperature drift and control the refractive power linearly. The refractive power ranges from -16 $m^{-1}$ to +17 $m^{-1}$ at 25°C and from -15 $m^{-1}$ to +28 $m^{-1}$ at 75°C. For the refractive power range at 25°C, the fluid pressure varies from - 270 Pa to + 270 Pa.

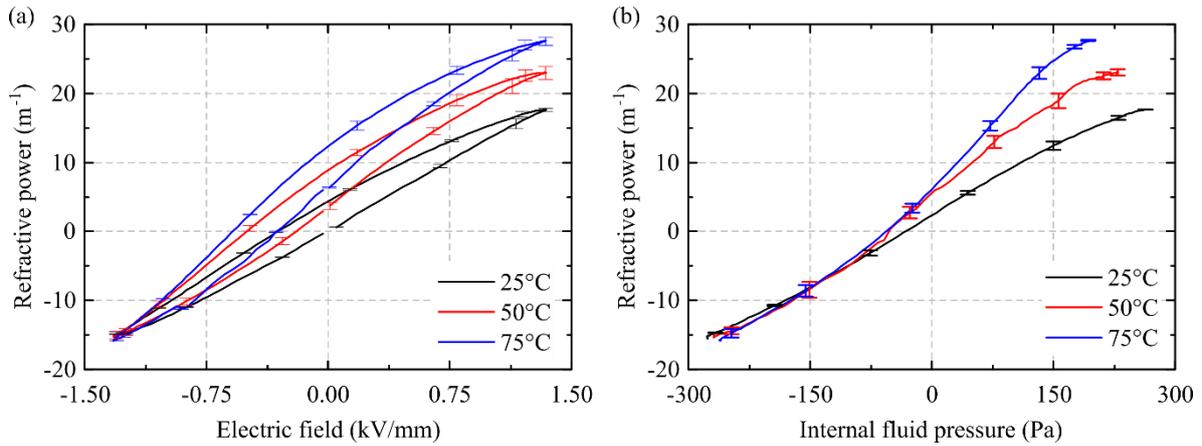

Figure 4 Temperature effect on the refractive power of the lens defined (a) as a function of the actuated voltage showing hysteresis response and (b) as a function of the internal fluid pressure.

The drift in refractive power seems stronger for positive than for negative electric fields; this can be explained by two effects: Firstly, the thermal expansion of the fluid results in an increased volume, which produces a positive membrane pre-deflection and hence a positive refractive power. Secondly, the increase in piezoelectric $d_{31}$ coefficient at higher temperatures [8] increases the net actuator deflection, which corresponds to higher negative and higher positive membrane deflections, thus higher positive and negative refractive powers. The superposed two effects result in a net higher positive refractive power.

## 4. SIMULATION

The asymmetric temperature drift of the refractive power observed in the experiment was underlined by a 2D axisymmetric model of the adaptive lens (Figure 5 (a)) simulated in *COMSOL Multiphysics®* [9]. The simulation model was used to analyze the combined effects of the fluid thermal expansion and the increased piezoelectric deflection at higher temperatures. Figure 5 (b) compares the simulation and experimental results that show a similar temperature drift of the refractive power. At low voltage, i.e., negative pressure, the temperature effect diminishes.

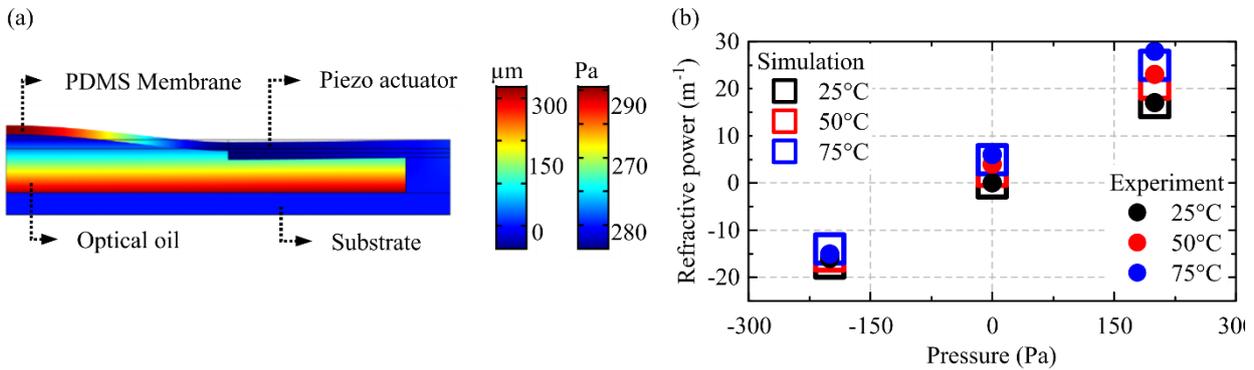

Figure 5 (a) The 2D axisymmetric model of the adaptive lens simulated in COMSOL Multiphysics® with the surface plot showing the deformation of the membrane and change in internal fluid pressure and (b) comparison of experiment and simulation results showing similar refractive power drift at a higher temperature.

## 5. CLOSED-LOOP CONTROL

The integrated sensors were used in a closed-loop arrangement (Figure 6 (a)), to define the refractive power as a function of the fluid pressure at different temperatures, shown in the continuous line plots in fig 5 (b). In contrast, with the knowledge of fluid pressure and temperature, any required refractive power in the range was able to be achieved, shown with the discrete dots in Figure 6 (b), which as well proves the repeatability and precise controllability of the lens.

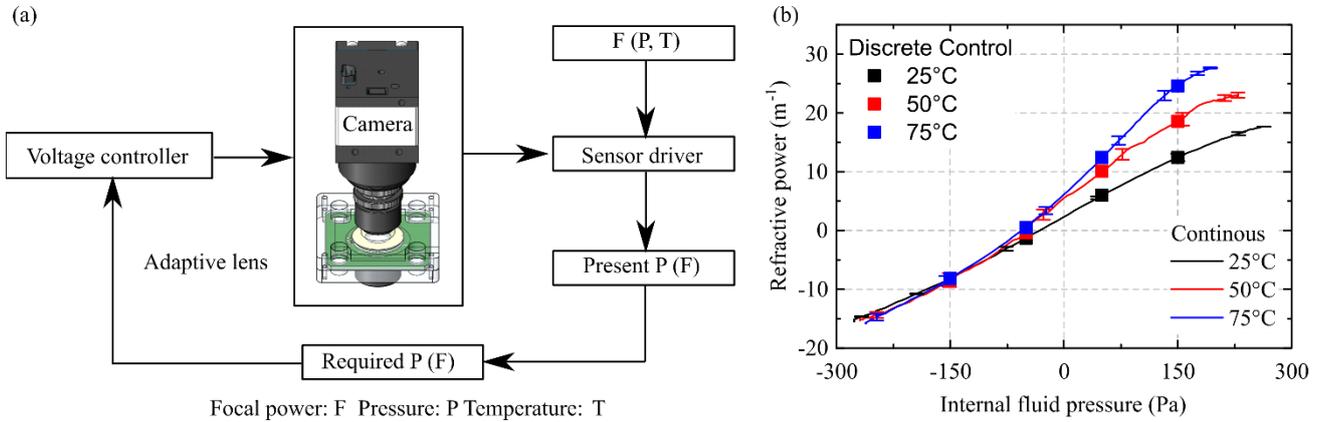

Figure 6 (a) Block diagram of the closed-loop system to control the refractive power and (b) continuous and discrete control of the refractive power defined as a function of fluid pressure and temperature.

## 6. SUMMARY

The adaptive lens with the integrated piezoelectric actuator and sensors allows for closed-loop control of the refractive power with a peak to peak range from 33 $m^{-1}$ at 25°C to 43 $m^{-1}$ at 75°C. The integrated pressure and temperature sensor address the non-linear hysteresis and thermal expansion effects to control the refractive power linearly. The increase in refractive power range at higher temperatures is due to the increased actuator deflection and the thermal expansion of the fluid.

**Acknowledgment**
This work was financed by the Baden-Württemberg Stiftung gGmbH under the project VISIR[2] (Variable Intelligent Sensors, Integrated and Robust for Visible and IR light).